\documentclass[11pt]{amsart}
\usepackage{latexsym}
\usepackage{amssymb}
\usepackage{amsxtra}
\usepackage{amsmath}
\usepackage{amsfonts}

\newcommand{\cleqn}{\setcounter{equation}{0}}
\newcommand{\clth}{\setcounter{theorem}{0}} 
\newcommand {\sectionnew}[1]{\section{#1}\cleqn\clth}
\newcommand{\beq}{\begin{equation}}
\newcommand{\eeq}{\end{equation}}
\newcommand{\beqa}{\begin{eqnarray}}
\newcommand{\eeqa}{\end{eqnarray}}
\newcommand{\beaa}{\begin{eqnarray*}}
\newcommand{\ben}{\begin{eqnarray*}}
\newcommand{\eaa}{\end{eqnarray*}}
\newcommand{\een}{\end{eqnarray*}}

\newcommand \nc {\newcommand}

\newtheorem{theorem}{Theorem}[section]
\newtheorem{lemma}[theorem]{Lemma}

\newtheorem{corollary}[theorem]{Corollary}

\newtheorem*{lemmaa}{Lemma A}
\newtheorem*{lemmab}{Lemma B}

\nc \thref{Theorem \ref}
\nc \leref{Lemma \ref}
\nc \prref{Proposition \ref}
\nc \coref{Corollary \ref}
\nc \deref{Definition \ref}
\nc \exref{Example \ref}
\nc \reref{Remark \ref}

\newcommand{\A}{\mathcal{A}}
\newcommand{\B}{\mathcal{B}}
\newcommand{\C}{\mathbb{C}}
\newcommand{\D}{\mathcal{D}}

\newcommand{\F}{\mathcal{F}}
\renewcommand{\H}{\mathcal{H}}
\newcommand{\J}{\mathcal{J}}
\renewcommand{\L}{\mathcal{L}}
\newcommand{\M}{\mathcal{M}}

\newcommand{\T}{\mathcal{T}}

\newcommand{\Z}{\mathbb{Z}}

\newcommand{\f}{\mathbf{f}}

\newcommand{\q}{\mathbf{q}}
\renewcommand{\t}{\mathbf{t}}
\newcommand{\x}{\mathbf{x}}
\newcommand{\y}{\mathbf{y}}

\def\Re{\mathop{\rm Re}\nolimits}
\def\res{\mathop{\rm Res}\nolimits}

\def\dim{\mathop{\rm dim}\nolimits}

\def\d{\partial}
\def\ev{\mathop{\rm ev}\nolimits}

\def\tensor{\otimes}

\def\Poincare{Poincar\'e}

\def\({\left(}
\def\){\right)}
\def\[{\left[}
\def\]{\right]}
\def\<{\left\langle}
\def\>{\right\rangle}

\def\lieGL{{\rm GL}}

\def\gl{\lambda}
\def\ge{\epsilon}
\def\ga{\alpha}

\def\gb{\beta}
\begin{document}
\title{ The Equivariant Gromov--Witten Theory of 
$\C P^1$ and integrable hierarchies}
\author{
Todor E. Milanov}
\thanks{E-mail: milanov@math.stanford.edu}
\date{}

\begin{abstract}
We construct an integrable hierarchy in terms of vertex operators and 
Hirota Quadratic Equations (HQE shortly) and we show that the equivariant 
total descendant potential of $\C P^1$ satisfies the HQE. Our prove is based
on the quantization formalism developed in \cite{G1}, \cite{G2}, and on the 
equivariant mirror model of $\C P^1.$ The vertex operators in our construction
obey certain transformation law under change of coordinates, which might be 
important for generalizing the HQE to other manifolds.   

We also show that under certain change of the variables, which is due to 
E. Getzler, the HQE are transformed into the HQE of the 2-Toda hierarchy.
Thus we obtain a new proof of the equivariant Toda conjecture. 

\end{abstract} 

\maketitle

\setcounter{section}{0}
\sectionnew{Introduction}

Let $\overline \M_{g,n}(\C P^1,d)$ be the moduli space of stable maps 
$f:(\Sigma,p_1,\ldots,p_n)\rightarrow \C P^1$, such that $\Sigma$ is a genus-$g$
complex curve with at most nodal singularities, $p_1,\ldots, p_n$ are 
marked points on $\Sigma$, pairwise distinct and different from the nodes, 
and $f$ has degree $d$ 
(i.e. $f_*([\Sigma])=d\,[\C P^1]\in H_2(\C P^1;\Z)$). 
Let $\L_i$ be the line bundle on $\overline\M_{g,n}(\C P^1,d)$,  whose fiber at  
$(\Sigma,p_1,\ldots,p_n;f)$ is the cotangent line $T^*_{p_i}\Sigma$ at the 
$i-$th marked point. We equip the moduli space $\overline\M_{g,n}(\C P^1,d)$ with 
the action of the complex torus $T=(\C^*)^2$, which is induced from the 
standard diagonal action of $T$ on $\C P^1$.

The equivariant descendant Gromov--Witten invariants are defined by:
\ben
\langle 
t_1\psi^{k_1},\ldots,t_m\psi^{k_m} 
\rangle_{g,n,d}^T = 
\int_{[\overline\M_{g,n}(\C P^1,d)]^T} \wedge_{i=1}^m (\ev_i^*(t_i)\cup \psi_i^{k_i}),
\een
where $t_i\in H^*_T(\C P^1;\mathbb{Q})$ is an equivariant cohomology class, 
$\ev_i$ is the evaluation map at the
$i-$th marked point, $\psi_i$ is the equivariant first Chern class of $\L_i$,
and $[\overline \M_{g,n}(\C P^1,d)]^T$ is the equivariant virtual fundamental cycle. 
The equivariant total descendant potential is defined by:
\ben
\D(\t)=\exp\(\sum \ge^{2g-2}\frac{Q^d}{n!}
\langle \t(\psi),\ldots,\t(\psi)\rangle_{g,n,d}^T \),
\een
where $\t(z) = \sum_{k\geq 0} t_k z^k $ is a formal series with coefficients 
in the equivariant cohomology algebra $H=H_T^*(\C P^1,\C)$ and 
the summation is over all $g,n,d\geq 0.$ 

Let  $\H:=H((z^{- 1}))$ be the loop space, equipped with the symplectic form
\ben
\Omega(f,g):=\frac{1}{2\pi i}\oint (f(-z),g(z))dz\ ,\ \ f(z),g(z)\in \H,
\een
where $(\ ,\ )$ is the equivariant {\Poincare} pairing. Note that 
$\H=\H_+\oplus \H_-\, ,$ where $\H_+=H[z]$ and $\H_-=z^{-1}H[[z^{-1}]]$, 
is a polarization of $\H$, which may be used to identify $\H$ with the 
cotangent bundle $T^*\H_+.$ The functions 
on $\H_+$ which belong to the formal neighborhood of $-{\bf 1}z$, where 
${\bf 1}$ is the unity in $H$, form a vector space which is called 
{\em the Bosonic Fock space}. The total descendant potential $\D$ is
identified with a vector in this space via {\em the dilaton shift} 
$\t(z)=\q(z)+z,$ where $\q(z)=\sum q_k\,z^k\ \in \H_+.$
 
An element $f\in \H$ can be written as
\ben
f(z) = \sum_{k=0}^\infty (q_{k,0}\phi_0+q_{k,\infty}\phi_\infty)\,z^k +
                         (p_{k,0}\phi^0+p_{k,\infty}\phi^\infty)\,(-z)^{-1-k},
\een
where 
$\{\phi_0=(p-\nu_\infty)/(\nu_0-\nu_\infty),
 \phi_\infty=(p-\nu_0)/(\nu_\infty-\nu_0)\}$ is a basis of $H$ 
and $\{\phi^0,\phi^\infty\}$ is the dual basis with respect
to the equivariant {\Poincare} pairing, $p$ is the equivariant first
Chern class of the hyperplane bundle $O(1)$ and $\nu_0$ and $\nu_\infty$ are 
the restrictions of $p$ respectively to the fixed points $[1,0]$ and $[0,1].$
The coordinate functions $\{p_{k,i},q_{k,i}\}_{k\geq 0,\, i=0,\infty}$ form a Darboux 
coordinate system on $\H$, thus the following quantization rules:
\ben
\widehat{p}_{k,i}:=\ge\d/\d q_{k,i},\ \ \ 
\widehat{q}_{k,i}:=q_{k,i}/\ge,\ \ \ i=0,\,\infty,
\een 
define a representation of the Heisenberg Lie algebra generated by the linear
Hamiltonians on $\H$ on the Bosonic Fock space. 

We introduce the vertex operators 
\beqa\label{vop_chi}
\Gamma^{\pm\chi_0} & = & \exp\(
\mp\sum_{d\in \Z} \gl^d\frac{
\prod_{j=-\infty}^0 (\nu-jz)}{
\prod_{j=-\infty}^d (\nu-jz)}\, \phi_0\ \ \ \)\sphat \\ \label{vop_chibar}
\Gamma^{\pm\chi_\infty} & = & \exp\(
\mp\sum_{d\in \Z} \gl^d\frac{
\prod_{j=-\infty}^0 (-\nu-jz)}{
\prod_{j=-\infty}^d (-\nu-jz)}\, \phi_\infty\)\sphat \ \ \,
\eeqa
\footnote{In this formula $\pm\chi_0$ and $\pm\chi_\infty$ stand for the 
characters of the torus action on the tangent spaces $T_0\C P^1$ and 
$T_\infty\C P^1$ respectively}where $\nu=\nu_0-\nu_\infty$, 
each of the exponents in \eqref{vop_chi} and \eqref{vop_chibar} is identified 
with a vector $f\in \H$ by expanding the 
terms corresponding to $d\geq 0\ (d<0)$ into the powers of $z^{-1}$ ($z$), and 
$\sphat$ stands for the following quantization rule. 
Given a vector $f\in\H$ we define a differential operator $\widehat{f}$ by
quantizing the linear Hamiltonian $\Omega(\ ,f).$ Expressions like
$e^f,\ f\in\H$ are quantized by first decomposing $f=f_-+f_+$, where 
$f_+(f_-)$ is the projection of $f$ on $\H_+ (\H_-)$, and then setting
$\(e^ f\)\sphat = e^{\hat f_-}e^{\hat f_+}$.

\begin{theorem}\label{HQE_descendents}
The equivariant total descendent potential of $\C P^1$ satisfies the following HQE:
\beqa\notag
&&
\res_{\gl=\infty} 
\left(\gl^{n-m}\Gamma^{\chi_0}\tensor\Gamma^{-\chi_0} - (Q/\gl)^{n-m}
\Gamma^{-\chi_\infty}\tensor\Gamma^{\chi_\infty}\right)\\ 
&&
\label{HQE}
\( e^{(n+1)\widehat\phi_0+n\widehat\phi_\infty}\tensor 
e^{m\widehat\phi_0+(m+1)\widehat\phi_\infty}\) 
\(\D\tensor\D\)\ \frac{d\gl}{\gl} =0\ ,
\eeqa
where $m,n$ are arbitrary integers.
\end{theorem}
The proof of \thref{HQE_descendents}  will be given in section 4.
The Hirota quadratic equations \eqref{HQE} are interpreted as follows. 
Switch to new variables $\x$ and $\y$ via the substitutions: $\q'=\x+\y$, $\q''=\x-\y$.
The LHS of the HQE expands as a series in $\y$ with coefficients Laurent 
series in $\gl^{-1}$, whose coefficients are quadratic polynomials in $\D$ and its 
partial derivatives. The residue is defined as the coefficient in front of 
$\gl^{-1}$. 

Let $y_0,y_1,\ldots $ and $ \overline y_0, \overline y_1,\ldots $ be 
two sequences of time variables related to $q_{0,0},q_{1,0},\ldots$ and 
$q_{0,\infty}, q_{1,\infty},\ldots$ via an upper-triangular 
linear change defined by the following relations:
\beqa
\label{qy1}
\sum_{n\geq 0} (-w)^{-n-1}\d_{q_{n,0}} & = &
\sum_{k\geq 0} \frac{1}{\nu(\nu-w)\ldots(\nu-(k+1)w)}\, \d_{y_{k}}, \\
\label{qy2}
\sum_{n\geq 0} (-w)^{-n-1}\d_{q_{n,\infty}} & = & 
\sum_{k\geq 0} \frac{1}{(-\nu)(-\nu-w)\ldots(-\nu-(k+1)w)}\, \d_{\bar y_{k}}.
\eeqa

\begin{corollary}\label{2toda_conjecture}
The change \eqref{qy1}--\eqref{qy2} transforms the sequence of functions 
$$
\{Q^{n^2/2}e^{n\ge\d}\D\}_{n\in\Z}, 
$$
where $\d= \d_{q_{0,0}}+\d_{q_{0,\infty}},$ into a sequence of 
tau-functions of the 2-Toda hierarchy. 
\end{corollary}

\coref{2toda_conjecture} is the so called equivariant Toda conjecture. The flows of the 
2-Toda hierarchy can be presented in many different ways (see \cite{UT}). Respectively, 
the equivariant 
Toda conjecture can be stated in many different ways. The original formulation belongs 
to E. Getzler \cite{Ge}
\footnote{Actually E. Getzler describes the flows of an integrable hierarchy which is a
reduction of the 2-Toda hierarchy and he calls it the equivariant Toda lattice.}, 
who described the 2-Toda flows as two infinite sequences of
commuting derivations in a certain differential algebra.    

\coref{2toda_conjecture} was proven by A. Okounkov and R. Pandharipande in \cite{OP2}. 
In section 2 we will prove that the change \eqref{qy1}--\eqref{qy2} transforms the 
HQE \eqref{HQE} into the HQE of the 2-Toda hierarchy. In particular, 
\thref{HQE_descendents} can be 
derived from the results in \cite{OP2}. However, we want to give an alternative proof
based on the equivariant mirror model of $\C P^1$. Our hope is that the argument 
can be genralized to $\C P^n$ and other toric manifolds.

\medskip

Let $\pi:\overline \M_{g,n+l}(\C P^1,d)\rightarrow \overline{\M}_{g,n}$ be the 
stabilization of 
the map which forgets the stable map and the last $l$ marked points. 
The {\em ancestor} Gromov--Witten invariants are defined by
\ben
\langle t_1\,\overline\psi^{k_1},\ldots,t_n\,\overline\psi^{k_n} 
\rangle_{g,n}^T(\tau)
=\sum_{d,\,l} \frac{Q^d}{l!}\int_{[\overline \M_{g,n+l}(\C P^1,d)]^T}
\wedge_{i=1}^n \( \ev_i^*(t_i)\overline\psi_i^{k_i}\) 
\wedge_{j=n+1}^{n+l} \ev_j^*(\tau),
\een
where $\tau,t_i\in H^*_T(\C P^1;\mathbb{Q})$, 
$\overline{\psi}_i=\pi^*(\psi_i)$ are the  pullbacks of the 
$\psi_i-$classes on $\overline{\M}_{g,n},$ and $Q$ is the Novikov variable. 
The total ancestor potential of $\C P^1$ is defined by:
\ben
\A_\tau(\t) = \exp\(\sum \ge^{2g-2}\frac{1}{n!}
\langle \t(\overline\psi)+\overline\psi,\ldots,
        \t(\overline\psi)+\overline\psi 
\rangle_{g,n}^T(\tau) \),
\een     
where $\t(z) = \sum t_k z^k,$ $t_k\in H^*_T(\C P^1;\mathbb{Q})$ and the 
summation is over all integers $g,n\geq 0$. It is a formal function 
in the sequence of vector variables $t_0,t_1,t_2, \ldots$ and $\tau.$
We identify $\A_\tau$ with a vector in the Fock space via the dilaton shift
$\t(z)=\q(z)+z.$

The total ancestor potential also satisfies HQE similar to \eqref{HQE}. 
Moreover, $\A_\tau$ has some special property which allows to interpret 
the corresponding residues as actual residues of meromorphic forms. 

An {\em asymptotical function} is, by definition, an expression 
\ben
\T = \exp \sum_{g=0}^\infty \ge^{2g-2}\T^{(g)}(\t;Q),
\een
where $\T^{(g)}$ are formal series in the sequence of vector 
variables $t_0,t_1,t_2,\ldots$  with coefficients in the Novikov
ring $\C[[Q]].$ Furthermore, $\T$ is called {\em tame} if
\ben
\left.\frac{\d}{\d t_{k_1,a_1}} \ldots \frac{\d}{\d t_{k_r,a_r}}\right|_{\t=0} 
\T^{(g)} = 0 \quad \mbox{whenever} \quad
k_1+k_2+\ldots +k_r > 3g-3+r,
\een
where $t_{k,a}$ are the coordinates of $t_k$ with respect to 
$\{\phi_0,\phi_\infty\}.$
The total ancestor potential $\A_\tau$ is a tame asymptotical 
function, because the tameness conditions is trivially satisfied
for dimensional reasons: $\dim \overline \M_{g,r}=3g-3+r.$ 

Introduce the vertex operators
\beqa\label{vop_chit}
\Gamma_\tau^{\pm\chi_0} = \exp ( \f_\tau^{\pm\chi_0}(x))\sphat=
\exp\( \pm \sum_{n\in \Z} I_{\chi_0}^{(n)}(\tau,x)\,(-z)^n \)\sphat,
\eeqa
where $I_{\chi_0}^{(n)}(\tau,x)$ are rational functions in $x$ with values in $H$, 
defined by  
\ben
I_{\chi_0}^{(0)}(\tau,x) = -\frac{\phi_0+(Qe^t/x^2)\,\phi_\infty}
{1-(\nu/x)-(Qe^t/x^2)}
\een
and the recursion 
\ben
\d_x I_{\chi_0}^{(n  )}(\tau,x) = \(1-\frac{\nu}{x}-\frac{Qe^t}{x^2}\)
     I_{\chi_0}^{(n+1)}(\tau,x)\ .
\een
Note that for $n<0$ the recursion involves a choice of integration constants.
In section 3, we will prove that the rationality of $I_{\chi_0}^{(n)}(\tau,x)$ 
with respect to $x$ uniquely determines the integration constants.

We say that: {\em a tame asymptotical function $\T$ satisfies the  
equivariant HQE of $\C P^1$, corresponding to a parameter $t\in \C$, if for each
$m\in \Z$:} 
\beqa\label{HQE_ancestor_potential}
\sum_{\xi=0,\infty}\res_{x=\xi} \ 
\exp 
\left\{2\frac{Qe^t}{x\nu}+\frac{m-1}{\nu}\(x+\frac{Qe^t}{x}-\nu\log x+t\nu_0\) 
\right\}\times \\
\notag
\times\left(\Gamma_\tau^{\chi_0}\tensor\Gamma_\tau^{-\chi_0}\right)
\(\T\tensor\T \)\frac{dx}{x^2}=0.
\eeqa
The Hirota quadratic equations \eqref{HQE_ancestor_potential} are interpreted as follows: 
switch to new variables $\x$ and $\y$ via the substitutions: 
$\q'=\x+\ge\y$, $\q''=\x-\ge\y$.
The first line in \eqref{HQE_ancestor_potential}
is independent of $\q'$ and $\q''$. Due to the tameness 
(\cite{G1}, section 8, Proposition 6), after cancelling the terms 
independent of $x$, the second line in \eqref{HQE_ancestor_potential}
expands into a power series in $\y$ and $\ge$, such that each coefficient 
depends polynomially on finitely many $I^{(n)}(\tau,x)$ and finitely 
many partial derivatives of $\A_\tau$. The residues in 
\eqref{HQE_ancestor_potential} are interpreted as the residues of 
meromorphic 1-forms.  
\begin{theorem}\label{HQE_ancestors2}
Let $\tau=t P.$
Then the total descendant potential of $\C P^1$ satisfies the HQE 
\eqref{HQE} iff 
the total ancestor potential $\A_\tau $ satisfies the HQE 
\eqref{HQE_ancestor_potential}.
\end{theorem}  
The proof of \thref{HQE_ancestors2} will be given in section 3.

{\bf Acknowledgments.} I am grateful to my adviser A. Givental for the 
many helpful discussions on this project and for showing me some computations
(see sections 2 and 3 bellow) which indicated that the equivariant Toda 
conjecture can be proved by applying the methods from \cite{G1} to the 
equivariant mirror model of $\C P^1.$

\sectionnew{Proof of \coref{2toda_conjecture}}

The goal in this section is to show that the upper-triangular linear change between 
the set of variables $\{y_k,\bar y_k\}$ and  $\{q_{n,0},q_{n,\infty}\}$ 
defined by \eqref{qy1}--\eqref{qy2} transforms the
HQE of the 2-Toda hierarchy into the HQE of \thref{HQE_descendents}
\footnote{
The observation that Getzler's change of the variables 
\eqref{qy1}--\eqref{qy2} transforms the vertex operators of 2-Toda into 
the vertex operators $\Gamma^{\pm\chi_0}$ and $\Gamma^{\pm\chi_\infty}$
belongs to A. Givental.}.

First, we apply the change \eqref{qy1} to the  terms in the exponent of 
\eqref{vop_chi} corresponding to $d=-k-1,$ $k\geq 0.$ 
According to the formula of E. Getzler (\cite{Ge}, Proposition A.1), 
inverting the change \eqref{qy1},
$$
\ge\d_{y_k} = (k+1) 
\sum_{i=1}^{k+1}\nu^{i}{k+1 \brack i}\ge\d_{k+1-i,0},
$$
where $\d_{n,0}:=\d/\d q_{n,0}$ and ${k\brack i},$ $1\leq i\leq k,$ is  
the coefficient in front of $\nu^iz^{k-i}$ in the 
polynomial: $\nu(\nu+z)\ldots(\nu+(k-1)z).$ 
Then 
\ben
\left(-\nu(\nu+z)\ldots(\nu + kz)\phi_0 \right)\sphat & = & \\
 =  
-\sum_{i=1}^{k+1}\nu^{i}{k+1 \brack i}\left( z^{k+1-i}\phi_0\right)\sphat &=&\\
  = 
-\sum_{i=1}^{k+1}\nu^{i}{k+1 \brack i}\ge\d_{k+1-i,0} &=& 
-(\ge\d_{y_k})/(k+1).
\een

Let
$y_k=\sum_{n\geq 0} a_{k,n}q_{n,0}$ be a linear change. Then by the chain
rule: $\d_{q_{n,0}} = \sum_{n\geq 0} \d_{y_k} a_{k,n}$ i.e.,
\beq\label{a_kn}
\sum_{n\geq 0} (-w)^{-n-1}\d_{q_{n,0}} = 
\sum_{k\geq 0} \(\sum_{n\geq 0} a_{k,n}(-w)^{-n-1} \) \d_{y_k}.
\eeq 
Using formula \eqref{a_kn}, we get that the change \eqref{qy1} transforms 
the terms in the exponent of \eqref{vop_chi} corresponding to $d=k+1,$ 
$k\geq 0$ as follows (note that $\phi^0=\nu\phi_0$): 
\ben
&&
\( -\frac{1}{(\nu-z)\ldots(\nu-(k+1)z)}\,\phi_0\)\sphat = \\
&&
\(-\sum_{n\geq 0} a_{k,n}(-z)^{-n-1}\phi^0\)\sphat=
\sum_{n\geq 0} a_{k,n}(q_{n,0}/\ge)=y_k/\ge,
\een
where we used that in the loop space formalism 
$q_{n,0}/\ge= (-(-z)^{-n-1}\phi^0)\sphat$. 

Finally, the term in the exponent of \eqref{vop_chi} corresponding to
$d=0$ is:  $-\widehat \phi_0 = -\ge \d_{0,0}.$
We get: $\Gamma^{\pm\chi_0}(\gl) = 
\Gamma^{\pm y}(\gl) e^{\mp\ge\d_{{0,0}}},$ where
\ben
\Gamma^{\pm y}(\gl)= 
\exp\(\pm\sum_{k\geq 0} ({y_k}/{\ge}) \gl^{k+1} \)
\exp\(\mp\sum_{k\geq 0}\ge\d_{y_k} \frac{\gl^{-k-1}}{k+1} \).
\een
In order to change the variables in $\Gamma^{\pm\chi_\infty}$ we need just to
switch index $0$ and index $\infty$. Note that 
$\nu=\nu_0-\nu_\infty$, thus when switching $0$ and $\infty$ we need to 
change $\nu$ to $-\nu$: 
$\Gamma^{\pm\chi_\infty}(\gl)
=\Gamma^{\pm\overline y}(\gl)e^{\mp\ge\d_{{0,\infty}}},$
where
\ben 
\Gamma^{\pm\overline y}(\gl) = 
\exp\(\pm\sum_{k\geq 0} (\overline y_k/\ge) \gl^{k+1} \)
\exp\(\mp\sum_{k\geq 0} \ge\d_{\overline y_k} \frac{\gl^{-k-1}}{k+1} \).  
\een 
Substituting the formulas for $\Gamma^{\pm\chi_0}$ and 
$\Gamma^{\pm\chi_\infty}$ in \eqref{HQE} and 
letting $\tau_n = Q^{n^2/2}e^{n\ge\d}\D$ we get:
\ben
\res_{\gl=\infty}\ \left\{ \gl^{n-m}\( \Gamma^y\tensor\Gamma^{-y}\)
\tau_n\tensor\tau_{m+1}  - 
\gl^{m-n}\( \Gamma^{-\overline y}\tensor\Gamma^{\overline y}\)
\tau_{n+1}\tensor\tau_{m}\right\}\frac{d\gl}{\gl}=0.
\een 
The last expression, up to rescaling the time variables $y_k$ and 
$\overline y_k$ by $\ge^{-k}$, is precisely the 
HQE of the 2-Toda hierarchy (see appendix A in \cite{O} or \cite{UT}).

\sectionnew{From descendants to ancestors}

\subsection{The twisted loop group formalism}
Let 
$$
\L^{(2)}\lieGL(H) = \left\{
M(z)\in \lieGL(\H)\ |\ M^*(-z)M(z)=1\right\},
$$
where $*$ means the transposition with respect to the {\Poincare} pairing, 
be the twisted loop group. 
The elements of the twisted loop group of the type
$M=1+M_1z+M_2z^2+\ldots$ (respectively $M=1+M_1z^{-1}+M_2z^{-2}+\ldots)$
are called upper-triangular (respectively lower-triangular) linear 
transformations. 
They can be quantized as follows: write $M=\log A,$ then $A(z)$ is an
infinitesimal symplectic transformation. We define  
$\widehat M=\exp \hat A,$ where $A$ is identified with the 
quadratic Hamiltonian $\Omega(A\f,\f)/2$ and on the space of
quadratic Hamiltonians the quantization rule $\sphat\ $ is defined by:
\ben
(q_{k,\ga}q_{l,\gb})\sphat := \frac{q_{k,\ga}q_{l,\gb}}{\ge^2},\ \ 
(q_{k,\ga}p_{l,\gb})\sphat := q_{k,\ga}\frac{\d}{\d q_{l,\gb}},\ \ 
(p_{k,\ga}p_{l,\gb})\sphat := \ge^2\frac{\d^2}{\d q_{k,\ga}\d q_{l,\gb} }.
\een 
We remark that $\sphat\ $ defines only {\em a projective representation} of
the subgroups of lower-triangular and upper-triangular elements of 
$\L^{(2)}\lieGL(H)$ on the Fock space $\B.$ 

Let  $S_\tau = 1+S_1z^{-1} + S_2z^{-2}+\ldots $ be an 
operator series defined by:
\ben
(S_\tau\phi_i,\phi_j) = (\phi_i,\phi_j)+\sum_{k\geq 0}
\langle \phi_i\psi^k,\phi_j\rangle_{0,2}^T(\tau)\, z^{-k-1}.
\een 
It is a basic fact in quantum cohomology theory that $S_\tau$ is an  
element of $\L^{(2)}\lieGL(H)$
(see \cite{G3}, section 6 and the references there in). 
According to \cite{CG}, Appendix 2,
\beq\label{descendent_ancestor}
\D = e^{ {\rm F}^{(1)}(\tau)}\widehat{S}_\tau^{-1}\A_\tau,
\eeq
where ${\rm F^{(1)}} = \left. \F^{(1)}\right|_{t_0=\tau,t_1=t_2=\ldots =0}$
is the genus-1 no-descendants potential.

The proof of \thref{HQE_ancestors2} amounts to conjugating the vertex 
operators in the HQE \eqref{HQE} by $S_\tau.$ 
We will use the following formula
(\cite{G1}, formula (17)):
\beq
\label{S-conjugation}
\widehat S_\tau e^{\hat \f}\hat S_\tau^{-1}  = 
e^{W(   \f_+  ,  \f_+   )/2}
e^{      (S_\tau \f) \sphat                      },
\eeq 
where $+$ means truncating the terms corresponding to the 
negative powers of $z.$ The key ingredients in the computation are an 
explicit formula for $S_\tau$ and the fact that the series  
$S_\tau\f^{\chi_i}(\gl),i=0,\infty$ obey certain transformation law under the 
change of the coordinate $\gl.$

\subsection{Explicit formula for $S_\tau$} 
By definition, {\em the  J-function} of $\C P^1$ is the following 
$H$-valued series:
\ben
J(\tau,z) = {\bf 1} +\tau z^{-1} + \sum_{k=0}^\infty   
\langle \phi_a\psi^k\rangle_{0,1}^T(\tau)\phi^a z^{-2-k}.
\een
According to \cite{G4}, Theorem 9.5, $J(\tau,z)=J{\bf 1},$ where $J$ is 
a linear operator on $H$ defined by
\ben
J= \sum_{d=0}^\infty \frac{(Qe^t)^d e^{(tp+t_0)/z}}
{\prod_{j=1}^d (p-\nu_0+jz)(p-\nu_\infty+jz)}.
\een
Here, the cohomology class $p$ is identified with the linear operator acting 
on $H$ by equivariant multiplication by $p$ and $(t_0,t)$ are 
the coordinates of $\tau$ in the basis $\{{\bf 1},P\}$ i.e., 
$\tau=t_0{\bf 1}+t p.$ The product in the denominator, by definition, 
is 1 if $d=0.$

Let $\tau=\tau_0\phi_0+\tau_\infty\phi_\infty.$ Then 
$t=(\tau_0-\tau_\infty)/\nu$
and $t_0=(-\nu_\infty\tau_0 + \nu_0\tau_\infty)/\nu.$ Using the 
chain rule and $z\d_{t_0}J=J$ we get  $z\d_{\tau_k} J=\phi_k(z\d_t)J,$ 
where the cohomology classes $\phi_k,$ $k=0,\infty$ are viewed 
as polynomials in $p$ and $\phi_k(z\d_t)$ is the differential 
operator obtained from $\phi_k$ by substituting $p$ with $z\d_t.$ 

Comparing the definitions of $S_\tau$ and the $J$-function we get
\beq\label{S_explicit}
\phi_i(\tau,z):=S_\tau\,\phi_i = \sum_{k=0,\, \infty} \phi^k(z\d_t)
(J\phi_i,\phi_i)\phi_k\ ,
\eeq 
where we used that $(J{\bf 1},\phi_i)=(J\phi_i,\phi_i).$

\subsection{Vertex operators and change of the coordinate $\lambda$}
 
Let $$\f(\gl)=\sum I^{(n)}(\gl)(-z)^n\in \H$$ be a vector such that 
the coefficients $I^{(n)}$ are formal Laurent series in $\gl^{-1}$, 
satisfying the recursive relation:
\beq\label{recursion} 
\d_\gl I^{(n)}(\gl) = \phi(\gl)I^{(n+1)}(\gl), 
\eeq
where  
$\phi = 1+a_1\gl^{-1}+a_2\gl^{-1}+\ldots $ is some formal series 
with $a_1\neq 0$.

First, we will prove that if such $\f$ exists then it is uniquely 
determined from 
$I^{(0)}$ and $\phi.$ The uniqueness is equivalent to the following Lemma.
\begin{lemma}\label{integration_constants}
Assume that all non-negative  modes $I^{(k)}$, $k\geq 0$ of the series $\f$ 
are zero.
Then $\f$ must be zero as well.
\end{lemma}
\proof
Assume that $I^{(-k)}=0$. Then \eqref{recursion} implies
$\d_\gl I^{(-k-1)} = \phi(\gl)I^{(-k)} = 0$. Thus $I^{(-k-1)}=c$ is a constant.
Again, by  \eqref{recursion} we have
\ben
I^{(-k-2)} = \int \phi(\gl)I^{(-k-1)}(\gl)d\gl = c\gl + c\,a_1\log \gl + 
(\mbox{ lower order terms}).
\een
By definition, all modes are Laurent series $\Rightarrow$ $c\,a_1=0$ 
$\Rightarrow$ $c=0.$
\qed
\begin{lemma}\label{transformation}
Let $\f$ be as above and $\gl=\gl(x)=x+O(x^{-1})$ be a transformation 
between two coordinates  
$\gl$ and $x$.  Then  
\ben
\(\sum_{n\in\Z} I^{(n)}(x)(-z)^n\)\ =\ 
\(\sum_{n\in\Z} I^{(n)}(\gl)(-z)^n\)
\exp\(\frac{1}{z}\int_x^\gl \phi(\eta)d\eta\).
\een
\end{lemma}
\proof
Denote the RHS by $g$ and note that 
\ben
(-z\d_x)g = (\phi(\gl)\gl'(x) - \phi(\gl)\gl'(x) +\phi(x))g = \phi(x)g. 
\een
Comparing the coefficients in front of $z$ we find that 
the modes of $g$ satisfy the recursive
relation \eqref{recursion}. Thus, thanks to \leref{integration_constants}, 
it is enough to show 
that the coefficients in front of $z^0$ are equal. On the RHS 
the coefficient in front of $z^0$ is 
\beq\label{I_0}
\sum_{k\geq 0} \frac{\(\Phi(x)-\Phi(\gl)\)^k}{k!}I^{(k)}(\gl),
\eeq
where $\Phi(\eta)$ is an anti-derivative of $\phi(\eta).$  
Let us introduce an auxiliary
variable $\xi = \Phi(\gl)$. Then the recursive relation \eqref{recursion} 
takes on the form
$\d_\xi I^{(n)}(\Phi^{-1}(\xi)) = I^{(n+1)}(\Phi^{-1}(\xi))$ $\Rightarrow$ 
$I^{(n)}(\gl) = \d_\xi^n I^{(0)}(\Phi^{-1}(\xi)).$ Thus, using the 
Taylor's formula, we can write \eqref{I_0} as 
\ben
I^{(0)}\(  
\Phi^{-1}(
\xi +\Phi(x) -\Phi(\gl))\) = I^{(0)}(x) ,
\een
since $\xi = \Phi(\gl)$ by definition. The lemma is proved. 
\qed

\subsection{Conjugating vertex operators}
According to formula \eqref{S-conjugation}, in order to conjugate 
$\Gamma^{\chi_i},$ $i=0,\infty$ by $S_\tau$ we need to
compute  $S_\tau\f^{\chi_i}$ and the corresponding 
phase factor $W_i := W(\f^{\chi_i}_+,\f^{\chi_i}_+).$
 
Similarly to \eqref{vop_chit}, we define the vertex operators
\beqa\label{vop_chit_bar}
\Gamma_\tau^{\pm\chi_\infty} = 
\exp ( \f_\tau^{\pm\chi_\infty}(\overline x)) =
\exp\( \pm \sum_{n\in \Z} 
I_{\chi_\infty}^{(n)}(\tau,\overline x)\,(-z)^n \)\sphat,
\eeqa
where $I_{\chi_\infty}^{(n)}(\tau,\overline x)$ are rational functions in 
$\overline x$ with values in $H$, 
uniquely determined (see \leref{integration_constants}) from 
\ben
I_{\chi_\infty}^{(0)}(\tau,\overline x) = 
-\frac{\phi_\infty+(Qe^t/\overline x^2)\,\phi_0}
{1+(\nu/{\overline x})-(Qe^t/\overline x^2)}
\een
and the recursive relation 
\ben
\d_{\overline x} I_{\chi_\infty}^{(n  )}(\tau,\overline x) = 
\(1+\frac{\nu}{\overline x}-\frac{Qe^t}{\overline x^2}\)
     I_{\chi_\infty}^{(n+1)}(\tau,\overline x)\ .
\een

\begin{lemma}\label{Sf}
Let $\f^{\chi_0}$ and $\f^{\chi_\infty}$ be respectively the exponents 
in \eqref{vop_chi} and \eqref{vop_chibar}. Then
\ben
\f_\tau^{\chi_0}(x) = S_\tau \f^{\chi_0}(\gl) ,
\quad
\f_\tau^{\chi_\infty}(\overline x) = S_\tau \f^{\chi_\infty} (\gl), 
\een
where 
$x=\gl(1+a_1\gl^{-1}+\ldots)$ and $\overline 
 x=\gl(1+\overline a_1\gl^{-1}+\ldots)$
are the unique solutions near $\gl=\infty$ respectively of 
\beqa\label{lambda_x}
&&
\gl-\nu\log \gl = x-\nu\log x +\frac{Qe^t}{x} +t\nu_0 \quad \mbox{ and } \\
&&
\label{lambda_xbar}
\gl+\nu\log\gl = 
\overline x +\nu\log \overline x +\frac{Qe^t}{\overline x}+t\nu_\infty.
\eeqa
\end{lemma}
\proof
We will prove the statement only for $\f_\tau^{\chi_0}.$ 
The argument for $\f_\tau^{\chi_\infty}$ is similar. Also, to avoid 
cumbersome notations, let us put $\chi:=\chi_0.$

Recalling formula \eqref{S_explicit} with $i=0$, we get:
\beqa\label{phi_0}
S_\tau\phi_0 & =   e^{t\nu_0/z}& \left[ \ 
\phi_0+
\sum_{d\geq 1}\frac{(Qe^t)^d}{\ \ d!z^d\ \ \nu(\nu+z)\ldots(\nu+(d-1)z)}\,
\phi_0 + 
\right. \\ \notag 
&& 
\left. \ \ \ \ \ \ \ +
\sum_{d\geq 1}\frac{(Qe^t)^d}{(d-1)!z^{d-1}\nu(\nu+z)\ldots(\nu+dz)}\,
\phi_\infty \right].
\eeqa  
Note that $(-z\d_\gl+(\nu/\gl))\f^\chi = \f^\chi$. Thus for each $d\in\Z$ 
we have:
\beq\label{f_chi}
\f^\chi = -\sum_{n\in\Z} \(-z\d_\gl+\frac{\nu}{\gl}\)^n\gl^d
\frac{\prod_{j=-\infty}^0(\nu-jz)}{\prod_{j=-\infty}^d(\nu-jz)}\,\phi_0 .
\eeq
Formulas \eqref{phi_0} and \eqref{f_chi} imply
\footnote{Formula \eqref{f_chit}  was derived by A. Givental.}
\beqa\label{f_chit} 
S_\tau \f^\chi = -\sum_{n\in\Z}\(-z\d_\gl+\frac{\nu}{\gl}\)^n
e^{(Qe^t/\gl + t\nu_0)/z}\[\phi_0 + \frac{Qe^t}{\gl^2}\ \phi_\infty\].  
\eeqa
Commuting $e^{(Qe^t/\gl + t\nu_0)/ z }$ across $(-z\d_\gl +\nu/\gl)^n$ in 
\eqref{f_chit}, we get:
\beq\label{f_series}
 S_\tau \f^\chi = -e^{(Qe^t/\gl+t\nu_0)/ z}\sum_n 
\(-z\d_\gl+\frac{\nu}{\gl}+\frac{Qe^t}{\gl^2}\)^n
\[\phi_0+\frac{Qe^t}{\gl^2}\,\phi_\infty \].
\eeq
Let $\f$ be the sum on the RHS of \eqref{f_series}. Note that $\f$ 
satisfies a recursive relation of the type \eqref{recursion} with
$\phi = 1-(\nu/\gl)-(Qe^t/\gl^2).$ Also, if $x$ is related to $\gl$ via
the change \eqref{lambda_x} then   
$$
\int_x^\gl \phi(\eta)d\eta =  
\(\, \gl-\nu\log \gl+ Qe^t/\gl\,\) - 
\(\, x-\nu x + Qe^t/x \,\) = Qe^t/\gl+{t\nu_0}.
$$
Thus the lemma follows from \leref{transformation}.
\qed

\begin{lemma}\label{W_chi}
Let $W_i=W_\tau(\f^{\chi_i}_+,\f^{\chi_i}_+),$ $i=0,\infty.$
Then the following formulas hold:
\beqa\label{w_chi}&&
W_0 = C_0 +2\frac{Qe^t}{\nu x} +
\log\frac{\gl(\gl-\nu)}{x^2-\nu x-Qe^t}\\
&& \label{w_chibar}
W_{\infty} 
= C_\infty-2\frac{Qe^t}{\nu \overline x} +
\log\frac{\gl(\gl+\nu)}{\overline x^2+\nu \overline x-Qe^t}\ ,
\eeqa  
where 
$C_0=t\nu_0/\nu + Qe^t/\nu^2$ and  
$C_\infty = -t\nu_\infty/\nu + Qe^t/\nu^2.$  
\end{lemma}
\proof
It is enough to prove the first formula. The second one is 
derived by switching indexes $0$ and $\infty.$ As before,
let us put $\chi=\chi_0.$

Using that $\d_x I^{(k)} = (1-\nu/x-Qe^t/x^2)I^{(k+1)}$ we get
{\allowdisplaybreaks
\ben
\d_xW_0 & = &
\d_x W_\tau(\f_+^\chi,\f_+^\chi) = 
\d_x\ \sum_{k,l\geq 0} (W_{kl}I_\chi^{(l)},I_\chi^{(k)})(-1)^{k+l} = \\
& = &
-\sum_{k,l\geq 0} \[([W_{k,l-1}+W_{k-1,l}]I_\chi^{(l)},I_\chi^{(k)}) \]
\(1-\frac{\nu}{x}-\frac{Qe^t}{x^2}\)  (-1)^{k+l}= \\
& = &
-\sum_{k,l\geq 0} \[ (S_l(-1)^lI_\chi^{(l)},S_k(-1)^k I_\chi^{(k)}) -
                    ( I_\chi^{(0)}(\gl)   ,I_\chi^{(0)}(\gl)) \]
\(1-\frac{\nu}{x}-\frac{Qe^t}{x^2}\)= \\
& = & 
\[ -\(        I_\chi^{(0)}(\tau,x),I_\chi^{(0)}(\tau,x)    \)+ 
    \(        I_\chi^{(0)}(\gl)   ,I_\chi^{(0)}(\gl)       \)    \]
    \(1-\frac{\nu}{x}-\frac{Qe^t}{x^2}\),
\een }
where the last equality follows from \leref{Sf}.

After a direct computation we get
\beqa\label{integral_1}
&& \int\(        I_\chi^{(0)}(\tau,x),I_\chi^{(0)}(\tau,x)    \)
    \(        1-\frac{\nu}{x}-\frac{Qe^t}{x^2}             \)dx = \\ 
\notag
&&\quad\quad
\frac{1}{\nu}\(-\frac{Qe^t}{x} + x-\nu\log x + \nu \log (x^2-\nu x-Qe^t)\),\\
\label{integral_2}
&&
\int \( I_\chi^{(0)}(\gl)   ,I_\chi^{(0)}(\gl)   \)\(1-\frac{\nu}{\gl}\)d\gl =
\frac{1}{\nu}\(\gl+\nu \log (\gl-\nu)\).
\eeqa

In order to fix the integration constant, note that $\f_+^\chi =\phi_0$ for
$x=\infty.$ Thus 
$$
C_0 := W_\chi|_{x=\infty} = W_\tau(\phi_0,\phi_0) = 
\(W_{0,0}\phi_0,\phi_0\) = \(S_1\phi_0,\phi_0\)= t\nu_0/\nu + Qe^t/\nu^2,
$$
where the last equality is obtained from \eqref{phi_0}. The lemma follows.
\qed

\subsection{Proof of \thref{HQE_ancestors2}}
Using formula \eqref{S-conjugation} together with \leref{Sf} and 
\leref{W_chi} we find that the HQE  \eqref{HQE} are equivalent to:
\beqa\label{HQE_ancestors}
& \res_{\gl=\infty} & 
\frac{d\gl}{\gl}
\left\{
\gl^{n-m}e^{W_0}\Gamma_\tau^{\chi_0}\tensor\Gamma^{-\chi_0}_\tau -
(Q/\gl)^{n-m}e^{W_{\infty}}\Gamma_\tau^{-\chi_\infty}\tensor
                                  \Gamma_\tau^{\chi_\infty}
\right\} \\ \notag
&&
\(e^{(n+1)\hat\phi_0(\tau,z)+n\hat\phi_\infty(\tau,z)}\tensor
  e^{m\hat\phi_0(\tau,z)+(m+1)\hat\phi_\infty(\tau,z)} \) 
\(\A_\tau\tensor\A_\tau\) = 0.
\eeqa

On the other hand $\A_\tau$ is a tame asymptotical function. Thus, 
after the substitutions 
\ben
\ge\y = (\q'-\q'')/2\ \ \mbox{ and } \ \ {\x} = (\q'+\q'')/2,
\een
and the cancellation of the terms which do not depend on $\gl$, the 1-form
in \eqref{HQE_ancestors} becomes a formal series in $\y$ and $\ge$
with coefficients which depend polynomially on finitely many of the modes 
$I_{\chi_0}^{(k)}$ and $I_{\chi_\infty}^{(k)}$ and finitely many
partial derivatives of $\overline\F_\tau(\x):=\log \A_\tau.$ 
According to \leref{Sf}, after choosing  new (formal) coordinates 
$x$ and $\overline x$ in a neighborhood of $\gl=\infty$, the coefficients 
$I_{\chi_0}^{(k)}$ and $I_{\chi_\infty}^{(k)}$ become rational functions 
respectively in $x$ and $\overline x$. 
Thus the residue in \eqref{HQE_ancestors} can be interpreted as the 
residues of rational 1-forms, which appear as the coefficients in a 
formal series in $\x,\y$, and $\ge$.
Moreover, the action of the operator (note that it is independent of $\gl$)
\beq\label{cancelation_operator}
\exp(-(n+1)\hat\phi_0(\tau,z)-n\hat\phi_\infty(\tau,z) )\tensor
\exp(-m\hat\phi_0(\tau,z)-(m+1)\hat\phi_\infty(\tau,z) )
\eeq
on such series is well defined: it results in transforming the formal 
series into a formal 
series  with coefficients which have the form of a rational 1-form multiplied 
by the exponential of a rational function, thus the residue still makes sense. 
Since the quantization of linear Hamiltonians is a representation of
Lie algebras, we have the following commutation relation:
$e^{\hat f}e^{\hat g} = e^{\Omega(f,g)}e^{\hat g}e^{\hat f}$. The operator 
\eqref{cancelation_operator}, applied  to \eqref{HQE_ancestors}, will cancel
the term in the $(\ )-$brackets. However, the two terms in the 
$\{\ \}-$brackets will gain commutation factors which are exponentials 
of the following expressions:
\ben
&&
\Omega\(\ \  -(n+1)\phi_0(\tau,z)-n\phi_\infty(\tau,z) \ \ ,\ \   
\f_\tau^{\chi_0}   \) + \\
&
+&\Omega\(\ \ -m\phi_0(\tau,z)-(m+1)\phi_\infty(\tau,z)  \ \ ,\ \   
\f_\tau^{-\chi_0} \)= \\
&
-&\Omega\((n+1)\phi_0+n\phi_\infty,\f^{\chi_0}) - 
\Omega(m\phi_0+(m+1)\phi_\infty,\f^{-\chi_0}\) = (m-n-1)\frac{\gl}{\nu} 
\een  
and
\ben
&&
\Omega\(\ \ 
-(n+1)\phi_0(\tau,z)- n\phi_\infty(\tau,z)\ \ ,\ \ 
\f_\tau^{-\chi_\infty} \ \)+  \\
&
+&\Omega\(
- m\phi_0(\tau,z)-(m+1)\phi_\infty(\tau,z)\ \ ,\ \ 
\f_\tau^{\chi\infty} \ \)=\\
&
-&\Omega\((n+1)\phi_0+n\phi_\infty,\f^{-\chi\infty}) -
\Omega(m\phi_0+(m+1)\phi_\infty,\f^{\chi\infty} \ \)
=(m-n+1)\frac{\gl}{\nu}. 
\een  
We get that the HQE \eqref{HQE_ancestors} are equivalent to:
\ben
& \res_{\gl=\infty} & 
\frac{d\gl}{\gl}
\left\{
\gl^{-m} e^{ W_0+(m-1)(\gl/\nu)  }
\Gamma_\tau^{\chi_0}\tensor\Gamma^{-\chi_0}_\tau - \right. \\
&&
\left.
-(Q/\gl)^{-m}e^{W_{\infty}\,+(m+1)(\gl/\nu)}
\Gamma_\tau^{-\chi_\infty}\tensor \Gamma_\tau^{ \chi_\infty}
\right\} 
 \(\A_\tau\tensor\A_\tau \)= 0,
\een

\medskip
Write the above residue sum  as a difference of two 
residues. In the first one substitute $\gl$ with $x$ according to the change 
\eqref{lambda_x} and use \leref{W_chi}:
\beqa
&&\label{change_x} 
\res_{\gl=\infty} \gl^{-m}e^{W_0+(m-1)\gl/\nu}
\(\Gamma_\tau^{\chi_0}\tensor\Gamma_\tau^{-\chi_0}\)
\(\A_\tau\tensor\A_\tau \)\frac{d\gl}{\gl} =\\
\notag
&&
\res_{x=\infty}\exp\(
2\frac{Qe^t}{x\nu}+\frac{m-1}{\nu}\(x+\frac{Qe^t}{x}-\nu\log x +t\nu_0\) 
+C_0 \) \times \\ \notag
&&
\times
\(\Gamma_\tau^{\chi_0}\tensor\Gamma_\tau^{-\chi_0}\)
\(\A_\tau\tensor\A_\tau \) \frac{d x}{x^2} .
\eeqa
Similarly, substituting in the second residue term $\gl$ with 
$\overline x$ we get:
\beqa
&&
\label{change_xbar} 
\res_{\gl=\infty} (Q/\gl)^{-m}e^{W_{\infty}\ +(m+1)\gl/\nu}
\(\Gamma_\tau^{-\chi_\infty}\tensor\Gamma_\tau^{\chi_\infty}\)
\(\A_\tau\tensor\A_\tau \)
\frac{d\gl}{\gl} 
=\\ \notag
&&
\res_{\overline x=\infty}\exp\(
-2\frac{Qe^t}{\overline x\nu}+\frac{m+1}{\nu}\(\overline x+
\frac{Qe^t}{\overline x}
+\nu\log \overline x +t\nu_\infty\) +C_\infty \)\times \\
\notag
&&
\times
\(\Gamma_\tau^{-\chi_\infty}\tensor\Gamma_\tau^{\chi_\infty}\)
\(\A_\tau\tensor\A_\tau \)
\frac{d \overline x}{\overline x^2} .
\eeqa
Note that $I_{\chi_\infty}^{(0)}(\tau,Qe^t/x)= -I_{\chi_0}^{(0)}(\tau,x)$ and
\ben
(-z\d_x)\f_\tau^{\chi_\infty} (Qe^t/x)= \(1+\frac{\nu}{x}-\frac{Qe^t}{x^2}\)
\f_\tau^{\chi_\infty}(Qe^t/x).
\een
Hence the substitution $\overline x = Qe^t/x$ transforms 
$\Gamma_\tau^{\mp\chi_\infty}$ into $\Gamma_\tau^{\pm\chi_0}.$ 
The residue \eqref{change_xbar} equals
\beqa\notag
\res_{x=0} Q^{-m}\exp\( 
-2\frac{x}{\nu} +\frac{m+1}{\nu}\(
\frac{Qe^t}{x}+ x+\nu\log Q +t(\nu + \nu_\infty )-\nu\log x \)
+C_\infty \) \times \\
\notag
\(\Gamma_\tau^{\chi_0}\tensor\Gamma_\tau^{-\chi_0}\)
\(\A_\tau\tensor\A_\tau\)\(-\frac{dx}{Qe^t}\) = \\
\notag
=-\res_{x=0}\exp\(
\frac{m-1}{\nu}\(\frac{Qe^t}{x}+x-\nu\log x +t\nu_0\)+2\frac{Qe^t}{x\nu}
+2\frac{t\nu_0}{\nu} + C_\infty-t\)\times \\
\label{change_xbar1}
\(\Gamma_\tau^{\chi_0}\tensor\Gamma_\tau^{-\chi_0}\)
\(\A_\tau\tensor\A_\tau\) 
\frac{dx}{x^2}
\eeqa
The difference between the exponents in the exponential factors in 
\eqref{change_x} and \eqref{change_xbar1} is:
\ben
C_0-C_\infty - t(2\frac{\nu_0}{\nu} -1) =t\frac{\nu_0+\nu_\infty}{\nu} 
-t\frac{2\nu_0-(\nu_0-\nu_\infty)}{\nu} = 0.
\een 
The theorem is proved. \qed

\sectionnew{Vertex operators and the equivariant mirror model of $\C P^1$}

In this section we give a proof of \thref{HQE_descendents} by 
showing that the total ancestor potential $\A_\tau$ satisfies 
the HQE \eqref{HQE_ancestor_potential}. We follow the argument in \cite{G1}. 

\subsection{Ancestor potential and the equivariant mirror model} 
According to \cite{G2},
for a generic $\tau\in H$ (in particular, $\tau=t\,p$ is generic), there are 
a basis  $\{{\bf 1}_1,{\bf 1}_2\}$ of $H$
and a linear transformation $M(\tau)=M_+\, M_-\,\in \L^{(2)}GL(H)$, 
where $M_+$ and $M_-$ are respectively upper- and lower-triangular 
linear transformations in $\L^{(2)}GL(H)$ such that 
\beq\label{ancestor_kdv}
\A_\tau = \widehat M \( \D_{\rm pt}(\q^1)\D_{\rm pt}(\q^2)\),
\eeq
where  $\q^i$ are the coordinates of $\q$ with respect to the 
basis $\{{\bf 1}_1, {\bf 1}_2\}$,  $\D_{\rm pt}$ is the Witten-Kontsevich 
tau-function, and $ \widehat M =\widehat M_+\, \widehat M_-.$

The transformation $M$ can be described in terms of the genus-0 Gromov--Witten
invariants of $\C P^1$. However we are going to use an alternative 
description in terms of the equivariant mirror model of $\C P^1$.  
Let $f:\C^2\rightarrow \C$ be the multi-valued function defined by
\ben
f(X_0,X_\infty) = X_0+X_\infty+\nu_0\log X_0 +\nu_\infty\log X_\infty\ .
\een 
Denote by $f_t$ the restriction of $f$ to the hypersurface
$\{X_0X_\infty =Qe^t\}$ and introduce the oscillating integrals:
\ben
\J_{\Gamma_i} = \int_{\Gamma_i\subset\{X_0X_\infty=Qe^t\}}e^{f_t/z}
\frac{dX_0dX_\infty}{d(X_0X_\infty)}, \quad
J_{\Gamma_i} = \sum_{k=0,\infty}\phi^k(z\d_t)\J_{\Gamma_i}\phi_k,
\een
where the cycles $\Gamma_i$ are defined as follows. Let $X^i$ and 
$u^i,\ i=1,2$ be the critical points and the critical values of 
$f_t$ respectively. Choose a path $\gamma_i$ in $\C$, starting at
$u^i$, avoiding the other critical value, and approaching $\infty$ in such
a way that $\Re \gamma_i\rightarrow -\infty$. The function $f_t$ is a 
double covering  in a neighborhood of the critical point $X^i$, hence 
$f_t^{-1}(\Lambda)=\{X_+(\Lambda),X_-(\Lambda)\}$ for $\Lambda\in\gamma_i$
close to $u_i$. Using the homotopy lifting property, we extend $X_+(\Lambda)$
and $X_-(\Lambda)$ for all $\Lambda\in\gamma_i$. The cycle $\Gamma_i$ is 
the union of two branches $\Gamma_i^+$ and $\Gamma_i^-$, parametrized by
$X_+(\Lambda)$ and $X_-(\Lambda),\ \Lambda\in\gamma_i$, respectively, with
orientation from $X_-$ to $X_+$. 

According to \cite{G2}, the matrix 
\ben
J=\begin{bmatrix} \phi^0(z\d_t)\J_{\Gamma_1} & \phi^0(z\d_t)\J_{\Gamma_2} \\
\phi^\infty(z\d_t)\J_{\Gamma_1} & \phi^\infty(z\d_t)\J_{\Gamma_2}
\end{bmatrix}
\een 
is asymptotic, as $z\rightarrow 0,$ to the matrix of the linear operator
$M$ with respect to the bases $\{{\bf 1}_1, {\bf 1}_2\}$ and 
$\{\phi_0,\phi_\infty\}$ respectively in the domain and codomain of $M$. 

\subsection{Conjugation by $\widehat M$} 
We identify the mirror hypersurface $\{X_0X_\infty=Qe^t\}$ with the complex 
circle $\C^*\subset \C P^1$ by choosing the coordinate $x=X_\infty.$ 
Let $x^i,i=1,2$ be the critical points of 
$f_t =Qe^t/x + x -\nu\log x +t\nu_0 +\nu_0\log Q$. In this subsection
we show how to conjugate the vertex operator $\Gamma_\tau^\chi(x)$ by 
$\widehat M$ for $x$ close to $x^i.$ 

The function $f_\tau$ is a double covering for $x$ close to $x^i.$ Let 
$x_\pm(\Lambda)$ be the two branches of $f_t^{-1}$. In order to keep track of 
them, we pick a reference point $\Lambda_0\in\gamma_i$ , sufficiently 
close to $u_i$. 
The value of $x_\pm$ at $\Lambda_0$ is fixed by requiring that 
$x_\pm(\Lambda_0)\in \Gamma_i^\pm$ ($\gamma_i$ and $\Gamma_i^\pm$
are the same as in the definition of $\Gamma_i$). For arbitrary $\Lambda$
(close to $u_i$), the value $x_\pm(\Lambda)$
depends on the choice of a path connecting $\Lambda_0$ and $\Lambda$. 

The oscillating integral $J_{\Gamma_i}$ can be written as:
\ben
J_{\Gamma_i}=\int_{\Gamma_i}e^{f_t(x)/z}\(\phi_0+\frac{Qe^t}{x^2}\,\phi_\infty\)\, dx =
\int_{u^i}^{-\infty} e^{\Lambda/z}I_{\chi_+-\chi_-}^{(0)}(\tau,\Lambda)d\Lambda,
\een
where  
$$
I_{\chi_+-\chi_-}^{(0)}(\tau,\Lambda)
:=I_{\chi}^{(0)}(\tau,x_+(\Lambda))-I_{\chi}^{(0)}(\tau,x_-(\Lambda)).
$$ 
For any $\ga=c_1\chi_++c_2\chi_-$ we define 

$$
\Gamma_\tau^\ga(\Lambda) = \exp(\widehat\f_\tau^\ga(\Lambda))= 
\exp\(\sum_{n\in\Z} 
I_\ga^{(n)}(\tau,\Lambda)(-z)^n\)\sphat,
$$
where
\footnote{Here
$\chi_\pm=[x_\pm(\Lambda_0)]\in H^0(f_t^{-1}(\Lambda_0;\Z)$ are one point cycles and
the RHS can be viewed as a $0$-dimensional integral: $\int_{\ga(\Lambda)}I_\chi^{(n)}(\tau,x)$.}
$$
I_\ga^{(n)}(\tau,\Lambda)=
c_1I_\chi^{(n)}(\tau,x_+(\Lambda))+ c_2I_\chi^{(n)}(\tau,x_-(\Lambda)).
$$
\begin{lemmaa}
For $\Lambda$ near the critical value $u^i$, the following formula holds: 
\beq\label{2toda_kdv}
\Gamma_\tau^{\pm(\chi_+-\chi_-)/2}\widehat M = 
e^{(W_i+w_i)/2}\widehat M \ \Gamma^{\pm},
\eeq
where
\ben
&&
W_i = -\lim_{\ge\rightarrow 0} \int_\Lambda^{u^i+\ge}\left\{\(
I_{(\chi_+-\chi_-)/2}^{(0)}(\tau,\xi),I_{(\chi_+-\chi_-)/2}^{(0)}(\tau,\xi) \) -
\frac{1}{2(\xi-u^i)}\right\}d\xi ,\\
&&
w_i = -\int_{\Lambda-u^i}^\Lambda \frac{d\xi}{2\xi} , \\
&&
\Gamma^\pm = \exp\(\sum_{n\in\Z} (-z\d_\Lambda)^n \frac{{\bf 1}_i}{\pm\sqrt{ 2\Lambda}}\).
\een 
\end{lemmaa}
\proof This is  Theorem 3 from \cite{G1}. \qed

The integration path in the definition of $W_i$ is any path connecting $\Lambda$ and $u_i+\ge$ and 
$\ge\rightarrow 0$ in such a way that $u_i+\ge\rightarrow u_i$ along a straight segment.The integration
path in $w_i$ is the straight segment connecting $\Lambda - u_i$ and $\Lambda$. The branch of 
$\sqrt \Lambda$ in $\Gamma^\pm$ is determined by the straight segment between $\Lambda-u_i$ and 
$\Lambda$ and the branch of $\sqrt{\Lambda-u_i}$. The later one is determined by the expansion:
\ben
I_{(\chi_+-\chi_-)/2}^{(0)}(\tau,\Lambda) = 
\frac{1}{\sqrt{2(\Lambda-u^i)}}\, ({\bf 1}_i + O(\Lambda-u^i)).
\een  
\begin{lemmab}
The vertex operators $\Gamma_\tau^{\pm\chi_\pm}$ factor as follows:
\ben
\Gamma_\tau^{\chi_\pm} = e^{\pm K}\,
\Gamma_\tau^{(\chi_\pm+\chi_\mp)/2} \,
\Gamma_\tau^{\pm(\chi_\pm-\chi_\mp)/2}, \quad
\Gamma_\tau^{-\chi_\pm} = e^{\pm K}\,
\Gamma_\tau^{-(\chi_\pm+\chi_\mp)/2}\, 
\Gamma_\tau^{\mp(\chi_\pm-\chi_\mp)/2},
\een
where
\ben
K=-\int_\Lambda^{u^i}\(
I_{(\chi_+-\chi_-)/2}^{(0)}(\tau,\xi),I_{(\chi_++\chi_-)/2}^{(0)}(\tau,\xi) \)d\xi.
\een
\end{lemmab}
\proof This is Proposition 4 from \cite{G1}, section 7. \qed  

\subsection{ Proof of \thref{HQE_descendents}.} 
It is enough to show that the total ancestor
potential $\A_\tau$ satisfies the HQE \eqref{HQE_ancestor_potential}.
Comparing with 
\leref{Sf} we see that the only possible poles of the vertex operators
$\Gamma_\tau^{\pm\chi}$ are at $x=0,\infty,x^1,$ or $x^2$, where $x^i,\ i=1,2$ 
are the critical points of $f_t(x)$.  
Thus if we want to prove that the total ancestor potential $\A_\tau$ satisfies
the HQE \eqref{HQE_ancestor_potential}, it is enough to show that the residue 
at each critical point $x^i$ is zero. 

Let us restrict $x$ to a neighborhood of $x^i.$ Using the change 
$\Lambda=f_t(x)$, we  transform the residue at $x^i$ into a residue 
at the critical value $u^i$. We have the following general formula:
\ben
\res_{x=x^i} g(x)dx= \res_{\Lambda=u^i} 
\sum_{\pm}g(x_\pm(\Lambda))\frac{\d x_\pm}{\d\Lambda}(\Lambda)d\Lambda,
\een
where $g(x)$ is an arbitrary function meromorphic in a neighborhood of
$x^i.$ 

The residue at $x=x^i$ of the 1-form in \eqref{HQE_ancestor_potential} 
transforms as follows:
\beq\label{res_xi}
\res_{\Lambda=u_i}
\left\{   d\Lambda \sum_{\pm} 
\frac{e^{2Qe^t/(\nu x_\pm)}  }
     {x_\pm^2-\nu x_\pm -Qe^t}
\Gamma_\tau^{\chi_\pm}\tensor\Gamma_\tau^{-\chi_\pm}\A_\tau\tensor\A_\tau
\right\} 
e^{\frac{m-1}{\nu}\Lambda}Q^{-(m-1)\nu_0/\nu},
\eeq
We will prove that the 1-form in the $\{ \ \}$-brackets in \eqref{res_xi} is analytic in 
$\Lambda$. In particular this would imply that the residue \eqref{res_xi} is 0.

\medskip

Using Lemma A and Lemma B we get:
\ben
&&
\sum_\pm \frac{e^{2Qe^t/(\nu x_\pm)}}{x_\pm^2-\nu x_\pm -Qe^t}\, 
\Gamma_\tau^{\chi_\pm}\tensor\Gamma_\tau^{-\chi_\pm}\,\A_\tau\tensor \A_\tau d\Lambda 
=\Gamma_\tau^{(\chi_++\chi_-)/2}\tensor \Gamma_\tau^{-(\chi_++\chi_-)/2} \\
&&
\widehat M \tensor \widehat M
\left\{\sum_\pm c_\pm(\tau,\Lambda)\Gamma^\pm_{(i)}\tensor\Gamma^\mp_{(i)}
       \frac{d\Lambda}{\pm\sqrt{2\Lambda}}
\right\} 
\prod_{i=1,2}\D_{\rm pt}(\q^i)\tensor \prod_{i=1,2}\D_{\rm pt}(\q^i),
\een
where the index $i$ in $ \Gamma^\pm_{(i)}$ is just to emphasize that the vertex operator is 
acting on the $i$-th factor in the product $\prod_{i=1,2}\D_{\rm pt}(\q^i)$ and the coefficients 
$c_\pm$ are given by the following formula:
\beq\label{c_pm}
\log c_\pm  = 
2\frac{Qe^t}{\nu x_\pm} - \log (x_\pm-\nu x_\pm-Qe^t)+W_i+w_i\pm 2K + \int_{\gamma_\pm}\frac{d\xi}{2\xi},
\eeq
where $\gamma_+$ and $\gamma_-$ are two paths connecting $1$ and $\Lambda$ and such  that 
$\int_{\gamma_\pm}d\xi/(2\xi) = \pm\sqrt{2\Lambda}$ (note that 
$\gamma_-^{-1}\circ\gamma_+$ is a simple loop around $0$). 

We will prove that with respect to $\Lambda$ the functions $c_+$ and $c_-$ are analytic and  coincide 
in a neighborhood of $u^i$. This would finish the proof of the theorem because, according to 
A. Givental \cite{G1}, the 1-form
\ben
\sum_{\pm }\Gamma^\pm_{(i)}\tensor\Gamma^\mp_{(i)}
       \frac{d\Lambda}{(\pm\sqrt{2\Lambda})} \T\tensor\T
\een
is analytic in $\Lambda$ whenever $\T$ is a tau-function of the KdV hierarchy. Thanks to
the Konstevich's theorem \cite{Ko}, $\D_{\rm pt}$ is a tau-function of the KdV hierarchy, thus the
theorem follows. 

Recalling the proof of \leref{W_chi}, formula \eqref{integral_1}, we get:
\ben
2\frac{Qe^t}{\nu x_\pm} - \log (x_\pm^2-\nu x_\pm-Qe^t) =  -\int_{\Lambda_0}^\Lambda 
\(I_{\chi_\pm}^{(0)}(\tau,\xi), I_{\chi_\pm}^{(0)}(\tau,\xi)\) d\xi + \Lambda/\nu+C_\pm 
\een
where the constants $C_\pm$ are independent of $\Lambda$ (they depends only on $x_\pm(\Lambda_0)$) 
and satisfy 
\ben
C_+-C_-=-\oint_{\gamma'}\(I_{\chi_-}^{(0)}(\tau,\xi), I_{\chi_-}^{(0)}(\tau,\xi)\) d\xi ,
\een
where  $\gamma'$ is a small loop around $u^i$ starting and ending at $\Lambda_0$. Formula 
\eqref{c_pm} transforms into
\ben
&&
\log c_\pm = \lim_{\ge\rightarrow 0} \\
&&
\left\{
-\int_{\Lambda_0}^\Lambda \(I_{\chi_\pm}^{(0)}(\tau,\xi), I_{\chi_\pm}^{(0)}(\tau,\xi)\) d\xi 
-\int_{\Lambda}^{u^i+\ge} 
\(I_{(\chi_+-\chi_-)/2}^{(0)}(\tau,\xi), I_{(\chi_+-\chi_-)/2}^{(0)}(\tau,\xi)\) d\xi \mp \right.\\
&&
\mp 2\int_{\Lambda}^{u^i+\ge}
\(I_{(\chi_+-\chi_-)/2}^{(0)}(\tau,\xi), I_{(\chi_++\chi_-)/2}^{(0)}(\tau,\xi)\) d\xi + \\
&&
\left.
\int_\Lambda^{u^i+\ge} \frac{d\xi}{2(\xi-u^i)} - \int_{\Lambda-u^i}^\Lambda\frac{d\xi}{2\xi} + 
\int_{\gamma_\pm}\frac{d\xi}{2\xi} +\frac{1}{\nu}\Lambda + C_\pm \right\}\ .
\een
In the first integral put $\chi_\pm=(\chi_\pm+\chi_\mp)/2 + (\chi_\pm-\chi_\mp)/2$. 
After a simple computation we get:
\beqa\label{c_pm1}
\log c_\pm & = &
-\int_{\Lambda_0}^\Lambda 
\(I_{(\chi_\pm+\chi_\mp)/2}^{(0)}(\tau,\xi), I_{(\chi_\pm+\chi_\mp)/2}^{(0)}(\tau,\xi)\) d\xi +  \\
\notag
&&
\lim_{\ge\rightarrow 0} 
\left\{ 
-\int_{\Lambda_0}^{u^i+\ge} 
\(I_{(\chi_+-\chi_-)/2}^{(0)}(\tau,\xi), I_{(\chi_+-\chi_-)/2}^{(0)}(\tau,\xi)\) d\xi - \right.\\
\notag
&&\
- 2\int_{\Lambda_0}^{u^i+\ge}
\(I_{(\chi_\pm-\chi_\mp)/2}^{(0)}(\tau,\xi), I_{(\chi_\pm+\chi_\mp)/2}^{(0)}(\tau,\xi)\) d\xi + \\
\notag
&&\ 
\left.
\int_{\gamma_\pm'} \frac{d\xi}{2\xi}\right\} +\frac{1}{\nu}\Lambda + C_\pm\ ,
\eeqa
where $\gamma_\pm'$ is the composition of the paths: $\gamma_\pm$ -- starting at 1 and ending at 
$\Lambda$, the straight segment between $\Lambda$ and $\Lambda-u^i$ (see the definition of $w_i$),
and the path from $\Lambda-u^i$ to $\ge$ obtained by translating the path between $\Lambda$ and 
$u^i+\ge$ (see the definition of $W_i$). Note that  formula \eqref{c_pm1} can be written also as
\ben
\log c_\pm & = &
-\int_{u^i}^\Lambda 
\(I_{(\chi_\pm+\chi_\mp)/2}^{(0)}(\tau,\xi), I_{(\chi_\pm+\chi_\mp)/2}^{(0)}(\tau,\xi)\) d\xi +  \\
&&
\lim_{\ge\rightarrow 0} 
\left\{ 
-\int_{\Lambda_0}^{u^i+\ge} 
\(I_{(\chi_\pm}^{(0)}(\tau,\xi), I_{(\chi_\pm}^{(0)}(\tau,\xi)\) d\xi 
+\int_{\gamma_\pm'} \frac{d\xi}{2\xi}\right\} +\frac{1}{\nu}\Lambda + C_\pm\ ,
\een
The integral on the first line is analytic near $\Lambda=u_i$, because near $\Lambda=u^i$, 
the mode $I^{(0)}_{\chi_\pm}$ expands as a Laurent series in $\sqrt{(\Lambda-u^i)}$ with singular term
at most $1/\sqrt{(\Lambda-u^i)}$. However the analytical continuation around $\Lambda=u_i$ transforms
$I^{(0)}_{\chi_\pm}$ into $I^{(0)}_{\chi_\mp}$, hence $I^{(0)}_{\chi_\pm}+I^{(0)}_{\chi_\pm}$ must be
single-valued and in particular, it could not have singular terms. 
The limit on the second line is clearly independent of $\Lambda$. The analyticity of $c_\pm$ follows. 

Note that 
\ben
\log c_+-\log c_- = \lim_{\ge\rightarrow 0} \left\{
-\oint_{\gamma_\ge} \(  I_{\chi_-}^{(0)}(\tau,\xi),I_{\chi_-}^{(0)}(\tau,\xi)\)d\xi +
\oint_{\gamma_-^{-1}\circ\gamma_+} \frac{d\xi}{2\xi} \right\},
\een  
where $\gamma_\ge$ is a closed loop around $u^i$ starting and ending at $u^i+\ge$. 
The second integral is 
clearly $\pm\pi i$ (the sign depends on the orientation of the loop $\gamma_-^{-1}\circ\gamma_+$). 
To compute the first one, write $\chi_-=(\chi_--\chi_+)/2 +(\chi_-+\chi_+)/2$ and 
transform  the integrand into
\ben
\(  I_{(\chi_--\chi_+)/2}^{(0)},I_{(\chi_--\chi_+)/2}^{(0)}\) + 
2\(  I_{(\chi_--\chi_+)/2}^{(0)},I_{(\chi_-+\chi_+)/2}^{(0)}\)+
\(  I_{(\chi_-+\chi_+)/2}^{(0)},I_{(\chi_-+\chi_+)/2}^{(0)}\).
\een
The last term does not contribute to the integral because it is analytic in $\xi$. The middle one
could have singular terms of the type $\mbox{ (something analytic in $\xi$)}/\sqrt{\xi-u^i}$, 
however $\lim_{\ge\rightarrow 0}\oint_{\gamma_\ge}$ of such terms is $0$.  Finally, the first term 
has an expansion of the type
\ben
\(  I_{(\chi_--\chi_+)/2}^{(0)},I_{(\chi_--\chi_+)/2}^{(0)}\) = \frac{1}{2(\xi - u^i)} + O(\xi -u^i)
\een 
and so it contributes only $\pm\pi i$ to the integral. Thus
$(\log c_+ -\log c_-) $ is an integer multiple of $2\pi i$, which implies that $c_+=c_-$.
The theorem is proved.
\qed


\end{document}